\documentstyle[11pt,IAU207_pasp,twoside,epsf]{article}
\def\be{\begin{displaymath}}
\def\ee{\end{displaymath}}
\def\mas{\mbox{\,M$_\odot$}}

\def\edcomment#1{\iffalse\marginpar{\raggedright\sl#1\/}\else\relax\fi}
\reversemarginpar

\begin{document}
\title{NGC 3603 -- a Local Template for Massive Young Clusters}
\author{Bernhard R. Brandl} 
\affil{Center for Radio Astronomy and Space Research, Cornell
  University, Ithaca, NY 14853, USA} 
\author{Wolfgang  Brandner}
\affil{European Southern Observatory, Karl-Schwarzschild-Str. 2,
  D-85748 Garching bei M\"unchen, Germany}
\author{Frank Eisenhauer} 
\affil{Max-Planck-Institut f\"ur Extraterrestrische Physik,
  Giessenbachstra\ss e, D-85741 Garching, Germany}
\author{Anthony F. J. Moffat} 
\affil{D\'{e}partment de physique, Universit\'{e} de Montr\'{e}al,
  Montr\'{e}al, Canada}
\author{Francesco Palla}
\affil{Osservatorio Astrofisico di Arcetri, Largo E.Fermi 5, I-50125
  Firenze, Italy}
\author{Hans Zinnecker}
\affil{Astrophysikalisches Institut Potsdam, An der Sternwarte 16,
  D-14482 Potsdam, Germany}

\begin{abstract}
  We present a study of the 
  star cluster associated with the massive Galactic H{\small II} region NGC
  3603 based on near-IR broad-- and narrowband observations taken with
  ISAAC/VLT under excellent seeing conditions ($\le 0.4''$).  We discuss
  color-color diagrams and address the impact of the high UV flux on
  the disk evolution of the low-mass stars.
\end{abstract}

\section{Stellar Content and Ionized Gas around NGC~3603}
NGC~3603 is the most massive Galactic H\,II region and star cluster
that can be studied
at optical wavelengths. The OB stars contribute more than $2000
M_\odot$ to the cluster mass, and provide about 100
times more ionizing power than those of the Trapezium cluster 
(Moffat, Drissen \&
Shara 1994, Kennicutt 1984).  At a distance of only 6~kpc it is the
perfect laboratory to study starburst-like conditions on a
star-by-star basis, and hence the conditions that may lead to the
formation and disruption of massive stellar clusters.

Eisenhauer et al. (1998) have studied the stellar IMF in NGC~3603 down
to $\sim 1\mas$ using JHK adaptive optics observations. 
We have further studied the stellar content of NGC~3603 in Brandl et al.
(1999) and found that the core region is well populated in low-mass
stars down to at least $0.1\mas$ with ages comparable to that of the
high mass stars.  Figure~1a shows the result of a 45 minute deep
H-band image taken with ISAAC at the VLT in April 1999 under excellent
seeing conditions ($\le 0.\!''4$).  The faintest stars visible in this
image are about $H=21^m$, corresponding to $0.1\mas$ for an age of
1~Myr.

\begin{figure}
\plottwo{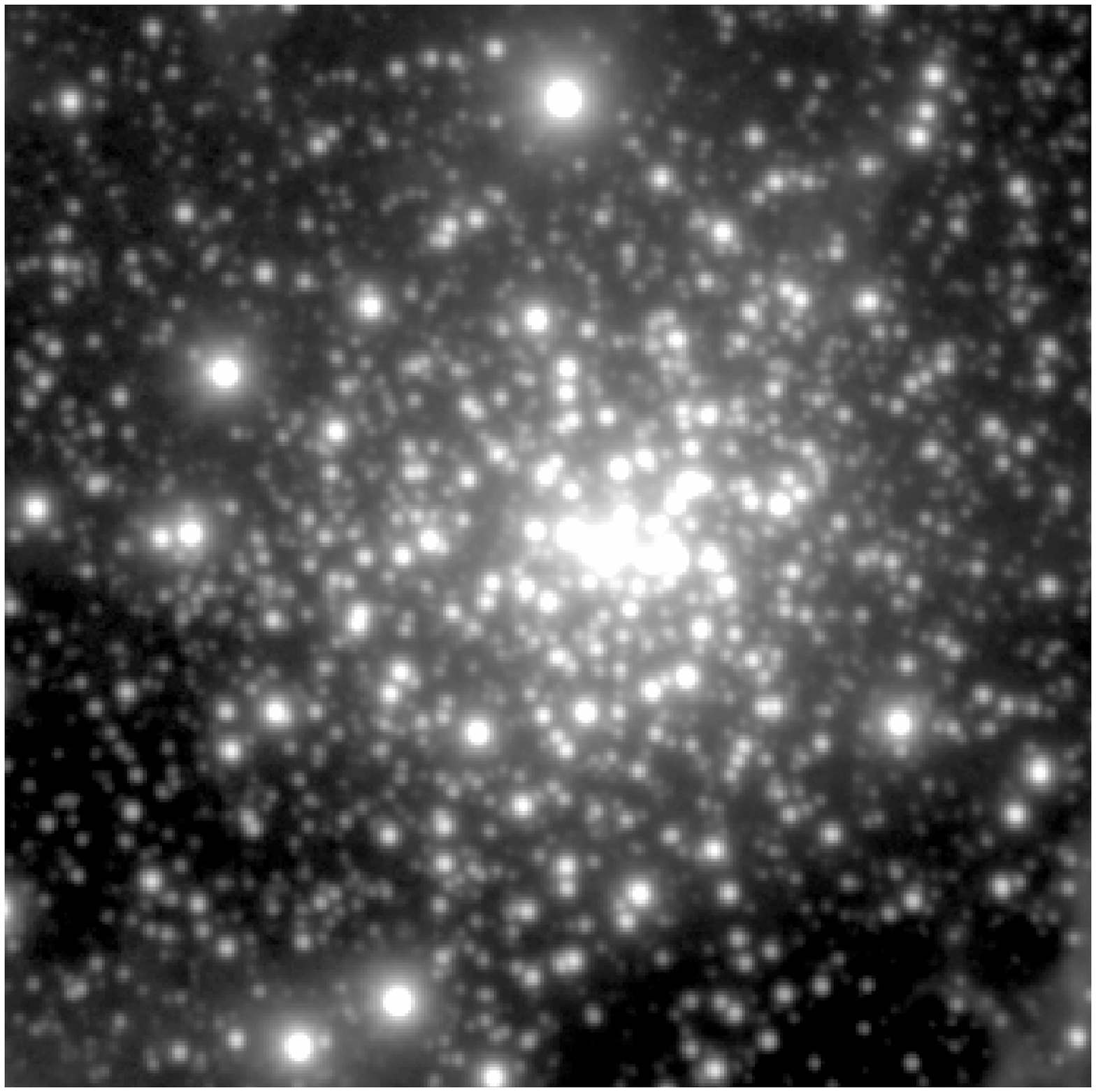}{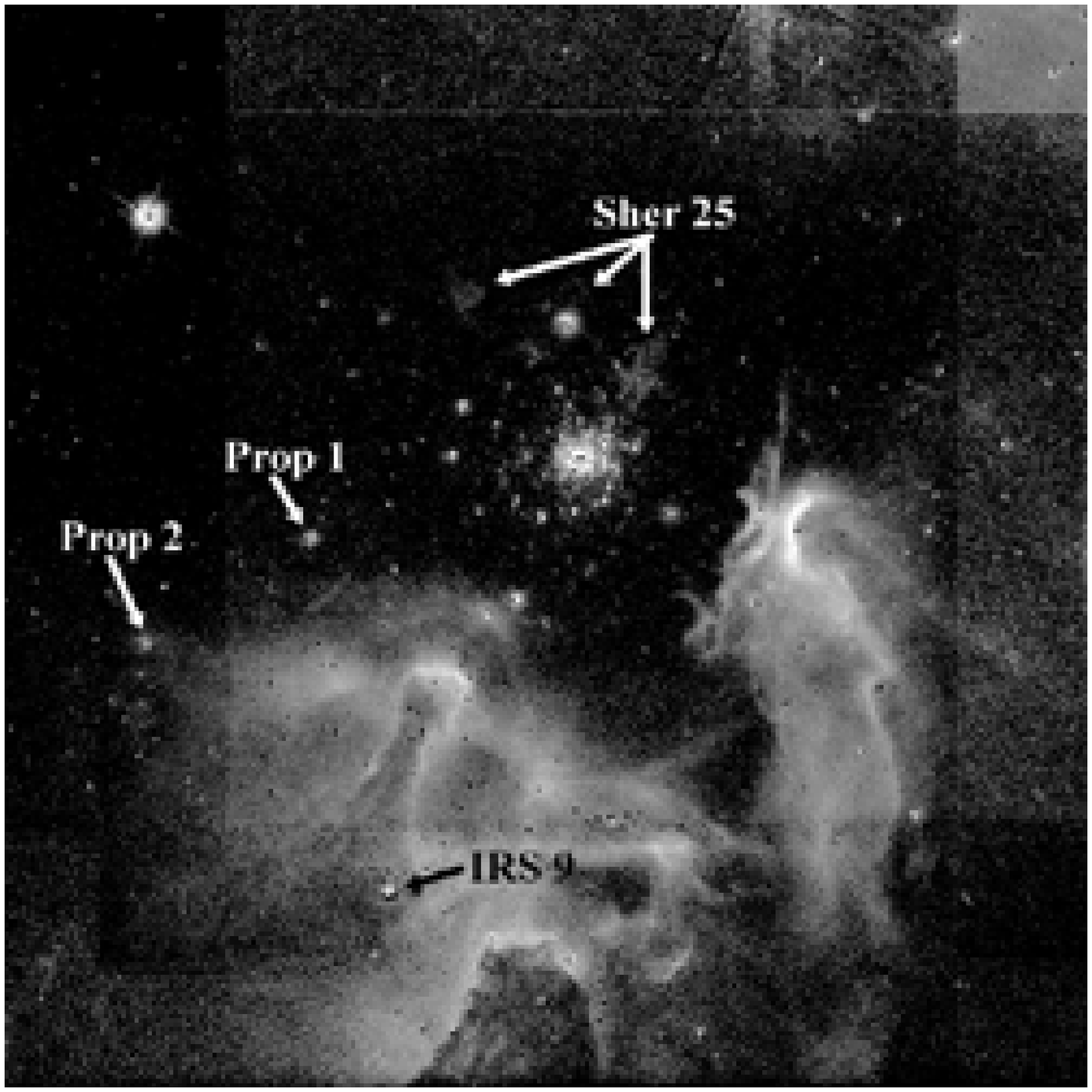}
\caption{\small \underline{Left:} Deep H-band ISAAC/VLT image of the 
  central $50''\times 50''$ ($1.55\times 1.55 \mbox{pc}^2$) of
  NGC~3603.  \underline{Right:} 150s exposure in the Bracket-$\gamma$
  line of the $2.\!'6\times 2.\!'6$ around NGC~3603; the continuum
  emission has been subtracted.}
\end{figure}

We have obtained ISAAC/VLT images in a variety of narrowband filters
(Br-$\gamma$, H$_2$, [FeII], HeI, CO-bandhead). Figure~1b shows a
short exposure in the Bracket-$\gamma$ line with a total integration
time of only 150 seconds; the K-band continuum emission has been
subtracted.  Despite the low signal-to-noise the extended nebular
structures and the ionization fronts to the west and southeast of
NGC~3603 are clearly visible.  The image reveals the ring nebula
around the blue supergiant Sher~25 (Brandner et al.  1997) about
$18''$ north of the core and its bipolar outflows to the east and
west.  The proplyd-like structures, recently discovered by Brandner et
al. (2000), are clearly visible.  The deeply embedded proto cluster
IRS~9 about $1'$ south of NGC~3603 becomes almost invisible after
continuum subtraction.

\begin{figure}[t]
\plotone{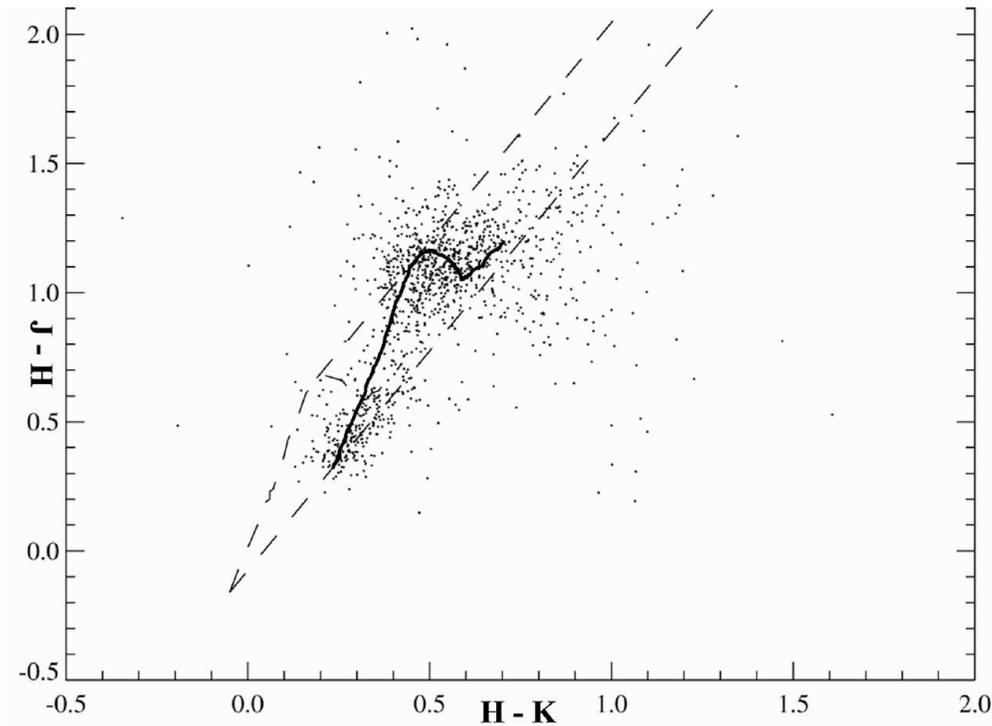}
\caption{\small (J-H)/(H-K) color-color diagram for all stars within 
  $r \le 33''$, where the field stars have been statistically
  subtracted.  The long dashed lines represent the reddening vectors.
  The thick line shows the location of (pre-) main sequence stars for
  $A_V = 4.5^m$.}
\label{testfields}
\end{figure}

Figure~2 shows the color-color-diagram for all stars within the
central parsec ($r \le 33''$) detected in all three JHK bands on the
same pixel.  To reduce the contamination from field stars we have
statistically subtracted the stellar density in the outer parts of our
$3.\!'4$ large FOV.  The solid line in Fig.~2 indicates the evolution
along the (pre-) main sequence for a foreground extinction of $A_V
= 4.5^m$.  The separation in two ``clumps'' indicates that only stars
more massive than about A5V ($\sim 2\mas$) have reached the main
sequence.

Note that classical T-Tauri or Herbig AeBe stars would fall in a
region to the right of the lower dashed line (extinction vector).
IR-excess emission from disks and circumstellar matter would make
their colors significantly different.  Despite the young age, there is
no evidence for a significant percentage of young stars with
circumstellar matter within the central parsec, probably because of
photoevaporation due to the high UV flux generated by the massive
stars.  If this result is generally true, no planetary systems should
be expected in globular clusters.

\acknowledgments
We'd like to thank Eva Grebel and Doug Geisler for organizing a
stimulating meeting in a beautiful location (Pucon).


\end{document}